\documentclass[a4paper,12pt]{article}

\usepackage{amsfonts}
\usepackage{verbatim}
\usepackage{latexsym}
\usepackage{epsfig}
\usepackage{graphicx}

\newcommand\ee{\end{eqnarray}}      %eqnarray
\newcommand\be{\begin{eqnarray}}

\def\l{\lambda}
\def\0{\nonumber}

\begin{document}

\begin{center}
{\Large \bf BRST, anti-BRST and their geometry}\\

\vskip 3cm

{\bf L. Bonora$^{(a, b)}$, R. P. Malik $^{(c, d)}$}  
\\{\it $^{(a)}$ International School for Advanced Studies (SISSA),}\\
{\it Strada Costiera, Via Beirut n.2-4, 34013 Trieste, Italy}\\
{\it $^{(b)}$ INFN, Sezione di Trieste, Trieste, Italy}\\
{\it $^{(c)}$ CAS, Physics Department, BHU, Varanasi-221 005, India }\\
{\it $^{(d)}$ DST-CIMS, Faculty of Science,  BHU, Varanasi-221 005, India} \\
 
{\small {\bf E-mails: bonora@he.sissa.it; malik@bhu.ac.in}}\\
\end{center}

\vskip 3cm

\noindent
{\bf Abstract:} \\

We continue the comparison between the field theoretical and geometrical 
approaches to the gauge field theories of various types, by deriving their 
Becchi-Rouet-Stora-Tyutin (BRST) and anti-BRST trasformation properties 
and comparing them with the geometrical properties of the
bundles and gerbes. In particular, we provide the geometrical interpretation 
of the so--called Curci-Ferrari conditions that are invoked for the absolute 
anticommutativity of the BRST and anti-BRST symmetry transformations
in the context of non-Abelian 1-form gauge theories
as well as Abelian gauge theory that incorporates a 2-form gauge field.
We also carry out the explicit construction of the 3-form gauge fields and 
compare it with the geometry of 2--gerbes.\\



\newpage

\section{Introduction}

In a gauge-fixed quantum gauge field theory, the local gauge symmetry is traded with the 
nilpotent BRST symmetry.
If the gauge-fixing is chosen in a particularly symmetric fashion, the BRST symmetry
is accompanied by a twin symmetry, called the anti-BRST. In this paper, we continue the
work started in \cite{BM} which consists in comparing the language of the 
quantum field theory with that of the underlying geometry.
In particular, in our present work, we concentrate primarily on two topics. The first one is a geometrical 
interpretation of the BRST and anti-BRST transformations in the twin cases of the non-Abelian
1-form gauge theories and theories incorporating the Abelian 2-form gauge potentials.
For the second topic, we focus on the Abelian 3-form gauge field 
theory and compare it with the 2-gerbe formalism. The above gauge theories,
as is well-known, are always endowed with the first-class constraints in the language
of Dirac's prescription for the classification scheme \cite{dirac} \cite{sund}.

As far as the first topic is concerned, we would like to understand, why in QFT, one needs the 
constraint relations (referred to as Curci-Ferrari (CF) relations) among the ghost fields 
and auxiliary fields in order to close the BRST and anti-BRST algebra.  
We start by reexamining an old problem: in the non-Abelian 1-form gauge field theories,
the requirement that the BRST and anti-BRST transformations must anti-commute, imposes a constraint 
on the (auxiliary) fields of the theory (the so-called CF condition). 
We will explain, first, how this fact can be given a geometrical interpretation.
This type of constraints, however, are characteristic, not
only of the non-Abelian 1-form gauge theories, but also of the higher
form Abelian (and perhaps non-Abelian) gauge theories whose field 
contents are based on the concept of gerbes. We wish to give a geometrical interpretation also of 
these constraints, which we keep referring to as the CF 
constraints.

Finally, as for our second topic,
we introduce a gauge theory based on the Abelian 3-form gauge potential and define its 
BRST and anti--BRST transformations. Furthermore, we obtain the CF-type conditions for their
absolute anti--commutativity. This is to be compared with the geometry of 2-gerbes. 
To this end we first introduce the latter, find their gauge transformations and show that
they can be reduced to the ones determined by the field theoretical methods. Thus, we establish
a connection between the field theory of the Abelian 3-form gauge theory with the geometry
associated with the 2-gerbes.

\section{An old problem: the CF constraints in non-Abelian gauge theories }

Let us consider a non-Abelian 1-form gauge theory with gauge group $G$. 
We write the gauge potential 1-form as $A= A_\mu^a(x) T^a dx^\mu$ where
$T^a$ are the anti-hermitian generators of $Lie (G)$. The appropriate
geometry for such a theory is well-known to be a principal
fiber bundle $P(M,G)$, the base space being the space-time manifold $M$ and with structure group $G$. A gauge 
transformation with infinitesimal parameter $\lambda= \lambda^a(x) T^a$, 
is given by
\be
\delta_\l A= D\lambda\equiv d \lambda+ [A,\lambda]. \label{gt}
\ee
Quantizing the theory requires gauge fixing, which, via the Faddeev-Popov
procedure, replaces the classical gauge invariance with the BRST invariance.
A BRST transformation is analogous to
a gauge transformation, except that the gauge parameter is replaced by an 
anticommuting field $\lambda \to c=c^a(x)T^a$. The origin of this 
transmutation was explained in \cite{BCR}: $c$ represents not a single gauge
transformation, but the whole set of gauge transformations. It is,
in fact, an alias of the Maurer--Cartan form $\omega$ {\it in the group of gauge 
transformations}. The fact that $\omega$ is a 1--form accounts for the
anticommutativity of $c$. Moreover, contracting it with left invariant 
vector fields in $P$ gives rise to all the gauge transformations. For this 
reason, we say that $c$ represents the whole set of gauge transformations.
In other words, the ghost field $c$ is the heuristic and compact form that QFT
adopts to express the geometric set-up of the non-Abelian 1-form gauge theories.

The BRST transformation (\ref{gt}) is not nilpotent unless we endow $c$ itself
with a BRST transformation. Since we have $s^2A=D(sc)+\frac 12 D([c,c])_+$,
the nilpotent BRST transformations must be as follows
\be
s A = D c, \quad\quad s c = -\frac 12 [c, c]_+. \label{BRST}
\ee
Nilpotency is simply a translation, in the language of
the QFT, of the fact that the gauge transformations form a Lie
algebra. In fact this implies that 
\be
(\delta_{\l_1}  \delta_{\l_2}-
\delta_{\l_2}  \delta_{\l_1})A- \delta_{[\l_1,\l_2]}A=0, \label{delta1delta2}
\ee and we see that 
(\ref{BRST}) exactly mimics this. In particular, the transformation
$sc$ mimics the last term at the LHS of (\ref{delta1delta2}).
This is important for the sequel: {\it nilpotency
of the BRST transformation is the quantum equivalent of the Lie algebra
(or Lie group) product law.} In other words, if the BRST transformations
were not nilpotent they could not be derived from a classical Lie group 
product law. 

If, in the quantization process, we use the Lorentz gauge fixing, 
then, the quantum theory turns out to be more symmetric. In this case, 
we have an additional symmetry, the anti-BRST symmetry. The parameter
of this new symmetry is the anti--ghost field $\bar c= \bar c^a(x)T^a$, which 
naturally appears in the gauge fixed action.

The complete set of BRST and anti-BRST transformations is 
\cite{CF,weinberg,HT}
\be
\matrix{ s A= Dc, & \bar s A =D\bar c, \cr
sc = -\frac 12 [c,c]_+, & \bar s\bar c = -\frac 12 [\bar c,\bar c]_+, \cr
s\bar c= B,& \bar s c= \bar B, \cr
s\bar B= [\bar B,c],& \bar s B=  [ B,\bar c],
}\label{bab}
\ee
the remaining transformations being trivial. The fields 
$B= B_\mu^a(x) T^a dx^\mu$ and $\bar B=\bar B_\mu^a(x) T^a dx^\mu$ are 
auxiliary scalar commuting fields. It is easy to see that $s$ and 
$\bar s$ are nilpotent. One expects the total
BRST-anti-BRST operator $s+\bar s$ to be nilpotent too. We have
\be
(s\bar s+\bar s s)A= D(B+\bar B+[c,\bar c]_+). \label{bab1}
\ee
Therefore nilpotency of $s+\bar s$ requires that
\be
B+\bar B+[c,\bar c]_+=0\label{CF}.
\ee 
This is the original CF constraint. On the other hand, we have seen 
above that nilpotency corresponds to Lie algebra structure. Therefore, it is
natural to require that (\ref{CF}) hold good. The nilpotency of $s+\bar s$ 
(which corresponds to anticommutativity) has
been used in the literature to study the physical spectrum of non-Abelian 1-form
gauge theories (see, e.g.  \cite{NO}). Added to it, the requirement of nilpotency of BRST and 
anti--BRST transformations is, since long, a standard folklore in the realm of BRST formalism.

Let us now delve into the geometrical interpretation of $\bar s$. The field
$\bar c$ is a sort of copy of $c$. Therefore, it would seem that the natural
geometrical setting for a BRST and anti-BRST-symmetric theory is 
$P(M,G\times G)$, that is a principal fiber bundle with structure group
$G\times G$, with $c$ taking values in the Lie algebra of the first group
and $\bar c$ in the Lie algebra of the second. But this cannot be the 
case since $[c,\bar c]\neq 0$.

We propose the following geometrical set-up. Starting from two isomorphic principal 
fiber bundles $P(M,G)$ and $Q(M,G')$ on the same base space, we can easily 
construct the bundle $(P+Q) (M,G\times G')$ on $M$ with structure group 
$G\times G'$, by pulling back the product $P(M,G)\times Q(M,G')$ via the
diagonal map $\Delta(x)= (x,x)$. The fibers of $P+Q$ are couples of 
fibers of $P$ and $Q$, i.e. they are $(p,q)$ such that $\pi_P(p)=x=\pi_Q(q)$.
The transition functions
$\hat\psi_{\alpha\beta}$ of $P+Q$ are given by
\be
\hat \psi_{\alpha\beta}\equiv(\psi_{\alpha\beta},\psi'_{\alpha\beta}): 
U_\alpha\cap U_\beta \to G\times G',
\ee
where
$\psi_{\alpha\beta}$ and $\psi'_{\alpha\beta}$ are the  transition functions
of $P$ and $Q$, respectively. 

Now, if $G\equiv G'$, in $\hat\psi_{\alpha\beta}$ we can choose, in 
particular, $\psi_{\alpha\beta}\equiv \psi'_{\alpha\beta}$. This means
that $P+Q$ is reducible to a bundle $R(M,G_d)$, with structure group
$G_d= Diag (G\times G)$ (see \cite{KN1}, Prop.5.3). It is obvious that
the fibers of $R$ are the diagonal fibers of $P+Q$. $R$ represents the
geometric set-up we are looking for. It is isomorphic
to $P(M,G)$, as it should be, but what matters is the fine structure of 
this isomorphism. In fact, it hosts two gauge groups, two copies 
of $G$, and this is what we need  to accommodate the above cited two different types of gauge transformations. 

Let us consider the relation between $R$ and $P+Q$. The reduction is specified 
by a homomorphism $R \to P+Q$, defined by a function $f$ such that  
$\forall u\in R$, $f(u)\in P+Q$ which reduces to the identity on the base 
$M$, and a group homomorphism $f_\delta : G\to Diag(G\times G)$, such
that $f(ug)= f(u) f_\delta(g)$ for any $u\in R$ and $g\in G$. A connection
${\cal A}$ in $P+Q$ reduces to a connection $A$ in $R$ and the two are related
by, \cite{KN1},
\be
f^*{\cal A}= f_\delta\cdot A, \label{redA}
\ee
where $f^*$ denotes the pull--back and 
$(f_\delta \cdot A)(X)= df_\delta(A(X))\equiv (A(X),A(X))$
for any vector field $X$ in $R$. The correspondence between
$A$ and ${\cal A}$ is one--to--one (see, e.g.  \cite{KN1} ch.II.6).

Let us consider now finite gauge transformations. They are given by vertical 
automorphisms $\psi$ of a principal bundle, i.e., by bundle morphisms 
that do not affect the basis. In the case of $R$, the fibers get transformed as
$\psi((p,p))=(\psi(p),\psi(p))$, where $\psi$ is an automorphism of $P=Q$. 
Now let us consider the same construction of $P+Q$ as above, but with $P$ 
replaced by $\psi_1^*P$, where $\psi_1$ is now an automorphism of $P$ (as opposed
to an automorphism of $Q$), i.e., we consider $\psi_1^*P+Q$ where 
$\psi_1^*P$ is the gauge transformed form of $P$. Naturally we would get
a version of $R$ with fibers $(\psi_1(p),q)$, where $q=\psi_1(q)$.
Therefore, trivially $\psi_1((p,q))=(\psi_1(p),q)$, which defines an
automorphism $\psi_1$ of $R$ originating from an automorphism of $P$. In the
same way, we can get automorphisms $\psi_2$ originated from the automorphisms of $Q$.

Taking a connection $A$ in $R$, we get, therefore, two types of gauge
transformations $\psi_1^*A$ and $\psi_2^*A$. We link them to
$c$ and $\bar c$, respectively. In this way $c$ ($\bar c$) is associated to the
Maurer-Cartan form of the first (second) factor in $G\times G$, respectively.
But they are both projected to the diagonal group. As a consequence, the anticommutator
$[c,\bar c]_+$ is non-vanishing. This seems to be the most appropriate geometrical
set-up for the BRST and anti-BRST.

Next, let us consider two infinitesimal gauge transformations of the first type 
($\psi_1$) and call them $\l_1$ and $\l_2$, and two of the second type
$\bar\l_1$ and $\bar\l_2$. It is easy to prove that
\be
\Big{(}(\delta_{\l_1}+\delta_{\bar\l_1})(\delta_{\l_2}+\delta_{\bar\l_2})-
(\delta_{\l_2}+\delta_{\bar\l_2})(\delta_{\l_1}+\delta_{\bar\l_1})\Big{)}A-
\delta_{[\l_1+\bar\l_1, \l_2+\bar\l_2]} A=0.\label{doubletransf}
\ee
This is the geometrical meaning of (\ref{bab}). However, in the anticommuting 
language, we cannot reproduce it without introducing the auxiliary fields $B$ and 
$\bar B$, for we have 
\be
(s\bar s+\bar s s)A= D( s \bar c+ \bar s c+[c,\bar c]_+),
\ee
which motivates the definition of $B$ and $\bar B$ in (\ref{bab}).

We stress again that (\ref{bab}) and (\ref{CF}) express, in the
language of quantum field theory, the simple Lie algebra rule 
(\ref{doubletransf}).  
 
Finally, let us make a comment concerning the Abelian case. When the 
gauge group $G$ is Abelian, we can, of course, repeat everything word by word.
In equations (3) and (4) the Lie brackets vanish, so BRST and anti--BRST
transformations are disconnected. The CF constraint is replaced by
$B+\bar B=0$.  These auxiliary fields are, in fact, decoupled and one can do
without them. We would like to
point out, however, that the 1-form Abelian bundle case is the only one in which
the CF constraint is superfluous. In the next more complicated case of the 
Abelian 2-form, which corresponds to the geometry of 1-gerbes, the CF 
constraint is essential.

\section{CF constraints for 1--gerbes}

The BRST and anti-BRST transformations for the 1-gerbes were worked out in 
\cite{BM}. Here we would like to give the relevant geometrical interpretation.

Let us recall some basic definitions. 
A 1--gerbe (for mathematical properties of gerbes see \cite{Giraud,Brylinski},
for previous applications in physics see \cite{gerbe}) may be characterized by 
a triple $(B,A,f)$, formed by
the 2-form $B$, 1-form $A$ and 0-form $f$, respectively. 
These are related in
the following way. Given a covering $\{U_i\}$ of 
the manifold $M$, we associate
to each $U_i$ a two--form $B_i$. On a double intersection $U_i
\cap U_j$, we have $B_i-B_j = dA_{ij}$. On the triple
intersections $U_i \cap U_j\cap U_k$, we must have $A_{ij}+
A_{jk}+A_{ki}= d f_{ijk}$. Finally, on the quadruple intersections
$U_i \cap U_j\cap U_k\cap U_l$, the following integral cocycle
condition must be satisfied:
\be
f_{ijl}-f_{ijk}+f_{jkl}-f_{ikl}= 2\;\pi \;n, 
\qquad n = 0, 1, 2, 3.........\label{cocycle} \ee
This integrality condition will not concern us in our Lagrangian
formulation but it has to be imposed as an external condition.

Two triples, represented by $(B,A,f)$ and $(B',A',f')$,
respectively, are gauge equivalent if they satisfy the 
following relations
\be
&&B_i'=B_i+ dC_i \quad\quad {\rm on} \quad U_i, \label{gaugeeq1}\\
&&A_{ij}' = A_{ij} + C_i-C_j + d\lambda_{ij} \quad\quad {\rm on}
\quad U_i\cap U_j, \label{gaugeeq2}\\ && f_{ijk}' = f_{ijk} +
\lambda_{ij}+\lambda_{ki}+ \lambda_{jk} \quad\quad {\rm on} \quad
U_i\cap U_j\cap U_k, \label{gaugeeq3} 
\ee 
for the one--forms $C$ and the zero--forms $\lambda$.

We will now define the BRST and anti--BRST transformations corresponding to these
geometrical transformations. As shown in \cite{BM}, one can proceed in two different ways.
Either one defines an action for the triple of local fields $(B,A,f)$, quantizes it by adding all
the ghost and auxiliary fields that are needed, and verifies that the quantum action has the two
BRST and anti--BRST symmetries below (this is what we do in section 4 for the 3--form gauge field). 
Or, more heuristically, by analogy with the 1--form gauge theory,
one starts from the (known) gauge transformations of the  $(B,A,f)$ fields and constructs 
the BRST and anti--BRST transformations by simply relying on nilpotency and consistency.
The two procedures lead to the same results up to minor ambiguities (see below).
It should be recalled that while the above geometric transformations are defined on
(multiple) neighbohrood overlaps, the BRST and anti--BRST transformations in quantum field theory  
are defined on a single local coordinate patch. These (local, field-dependent) transformations are the means QFT uses to record the underlying geometry.

The appropriate BRST and anti--BRST transformations are\footnote{As done in  \cite{BM},
we take into account here also the scalar field $f$.}
\begin{eqnarray}\matrix{
 s\, B = d C, & s\, A= C + d \lambda,&s\, f=\lambda+\mu,\cr
s \,C= -d \beta,  &    s\,\lambda = \beta, 
& s\,\mu=-\beta, \cr
 s \,\bar C = - K, & s\,\bar K = d \rho, & s\,\bar\mu=-g, \cr
 s \,\bar\beta = - \bar \rho,  & s\,\bar \lambda= g,  &
s\, \bar g= \rho, }\label{BRST2}
\end{eqnarray}
together with $s \; [\rho, \bar\rho, g, K_\mu, \beta] = 0$, and
\begin{eqnarray}
&& \bar s\, B = d\bar C, 
\qquad\bar s\, A = \bar C + d \bar \lambda,\qquad  \bar s\, f=\bar\lambda+\bar \mu,\nonumber\\
&&  \bar s\,\bar C = + d\bar \beta, \qquad
 \bar s \, \bar \lambda  = - \bar \beta, 
\qquad  \bar s\, \bar \mu=-\bar \beta, \nonumber\\
&& \bar s\, C = + \bar  K,
\qquad \bar s \, K = - d \bar \rho , \qquad \bar s \,\bar \mu =\bar g,  \nonumber\\
&& \bar s\,\beta = + \rho,    \qquad \bar s\,\lambda= - \bar g, \qquad
\bar s\, g = - \bar \rho, \label{antiBRST2}
\end{eqnarray}
while $\bar s \;[\bar\beta, \bar g, \bar K_\mu, \rho, \bar\rho]= 0$.

It can be easily verified that $(s+\bar s)^2=0$ if the following constraint
is satisfied:
\begin{equation}
\bar K- K=d\, \bar g -d\, g.\label{CF2}
\end{equation}
This condition is both BRST and anti--BRST invariant.
It is the analogue of the Curci-Ferrari condition in non--Abelian 1-form 
gauge theories and we will refer to it with the same name.

The field content and BRST and anti--BRST structure for 1--gerbe field theories
is shown schematically in Figure 1.

\begin{figure}[htbp]
    \hspace{-0.5cm}
\begin{center}
    \includegraphics[scale=0.5]{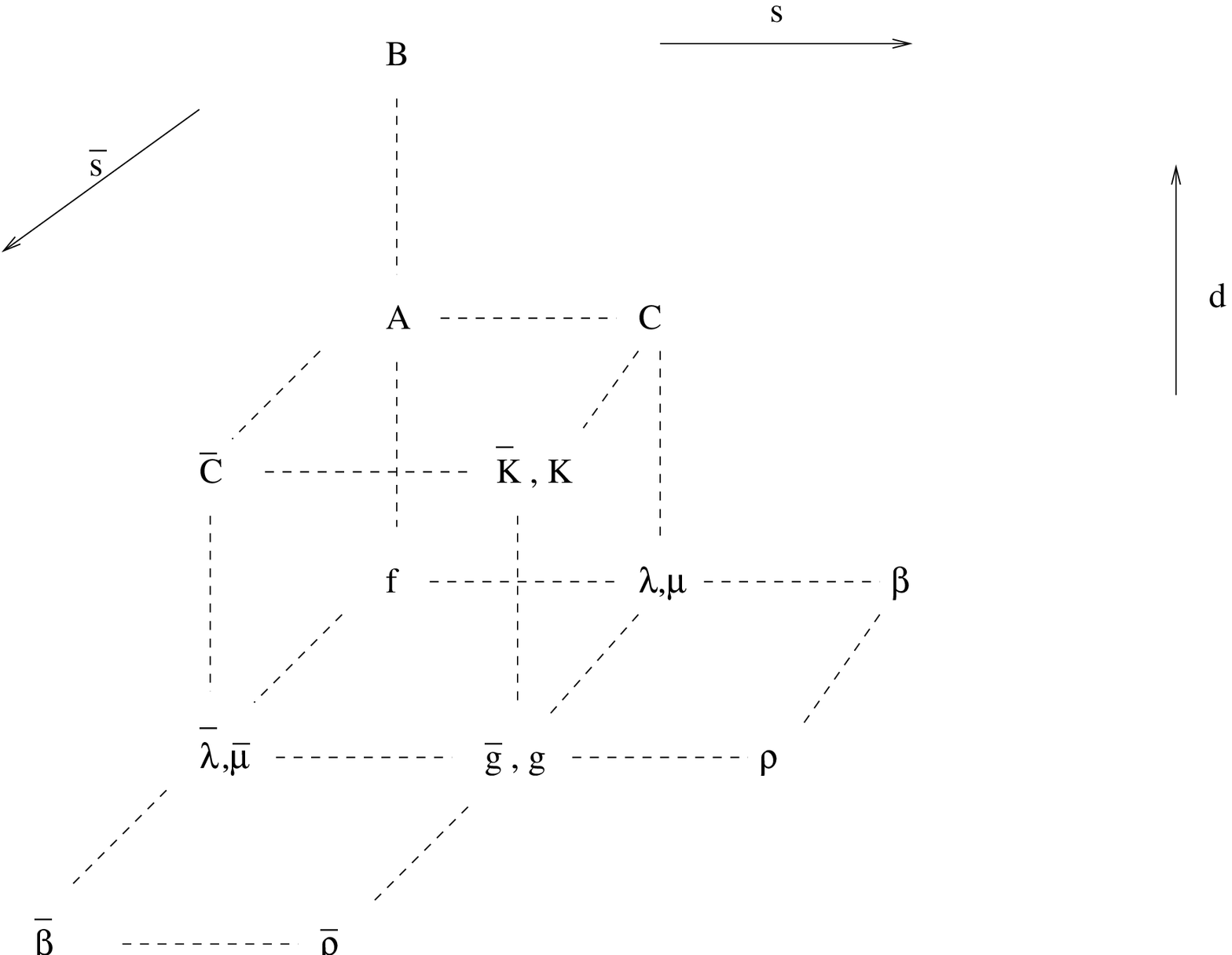}
    \end{center}
\caption{\emph{\small A schematic view of BRST and anti-BRST transformations
for 1--gerbes}}
    \label{fig:A}
\end{figure}

Before we proceed with the discussion, we would like to note that the 
above realization of the BRST and anti--BRST algebra is not the only 
possibility. In general, it may be possible to augment it by the 
addition of a sub-algebra of 
elements which are all in the kernel of both $s$ and $\bar s$ or,
if it contains such a sub-algebra, the latter could be moded out. 
For instance, in equations (\ref{BRST2},\ref{antiBRST2}) $\rho$ and $\bar \rho$
form an example of this type of subalgebra. It is easy to see that $\rho$ and 
$\bar \rho$ can be consistently set equal to 0.

We would like now to suggest a geometrical setting for these transformations and the
relevant CF constraints. We have already noted that for Abelian gauge 
bundles, we can carry out a construction similar to that of the previous 
section.  The Lie bracket terms in (\ref{bab}) are trivial and we can, in fact, 
dispense with the auxiliary fields $B$ and 
$\bar B$. That is to say the BRST and anti--BRST transformations are independent 
of any interference term and the CF condition is unnecessary. The gauge 
transformations, underlying a 1--gerbe, are of the Abelian type too, but, contrary to 
the 1-form gauge theory (i.e. 0-gerbe), they produce a non--trivial CF condition (\ref{CF2}).

The geometric set-up for the BRST and anti-BRST 1-gerbe transformations is
similar to the one that is true for the gauge bundles. 
One can define both Cartesian products and pullbacks of 1-gerbes,
(see, e.g. \cite{Brylinski}). We can, therefore, take the two copies of the same 1-gerbe
and make their product. Then, we pull the result back by the diagonal 
map (see above). So far, the construction is parallel to the one in the 
previous section. Now we have to do the analogue of the mapping to the
diagonal of $G\times G$. This is not so easy in the above formulation of the
1--gerbes. But there is another formulation due to Hitchin \cite{Hitchin} 
which we now recall and dwell upon. With respect to a covering $\{U_i\}$, a 1-gerbe 
is specified by the following data:
\begin{itemize}
\item a line bundle $L_{ij}$ for any intersection $U_i \cap U_j$,
\item an isomorphism between $L_{ij}$ and $L^{-1}_{ji}$,
\item a trivialization of $L_{ij}L_{jk}L_{ki}$, that is a map 
      $\theta_{ijk}: U_i \cap U_j\cap U_k \longrightarrow U(1)$,
\item this map satisfies the cocycle condition $\theta_{jkl} \theta^{-1}_{ikl}
\theta_{ijl} \theta^{-1}_{ijk}=1$.
\end{itemize}

In this definition, the line bundles are associated with the $U(1)$ principal bundles,
$L^{-1}$ represents the bundle dual to $L$, and the products of the bundles are 
tensor products. It is now possible to redo the construction of the previous 
section (including the diagonal mapping for fibers) locally, i.e., for each 
line bundle and each map $\theta$ (by replacing each local $P$ with a 
corresponding isomorphic $R$) while satisfying the conditions of the 
1--gerbe definition. This means that we can think of two copies of 
the transformations (\ref{gaugeeq1}--\ref{gaugeeq3}) 
with parameters $C_i,\lambda_{ij}$ and $\bar C_i,\bar\lambda_{ij}$, 
respectively. 

This is the classical set-up we propose in order to 
accommodate the BRST and anti--BRST 1--gerbe transformations.

From the two copies of (\ref{gaugeeq1}--\ref{gaugeeq3}),
it is easy to recognize the origin of many of the transformations in 
(\ref{BRST2},\ref{antiBRST2}). For instance from (\ref{gaugeeq1}), we
see that $C_i$ is not uniquely defined, we could choose $C'_i=C_i+d \beta_i$
(or $\bar C'_i=\bar C_i+d \bar\beta_i$), 
obtaining in this way, the classical analogue of $sC$ (or $\bar s\, \bar C$).
Plugging this into (\ref{gaugeeq2}), we get 
$\lambda'_{ij}=\lambda_{ij}-\beta_i+\beta_j$ and obtain the analogue of $s \,\lambda$,
etc. Denoting now, for simplicity, the classical
infinitesimal transformations $\delta$ and $\bar\delta$ 
corresponding to the two copies of 
(\ref{gaugeeq1}--\ref{gaugeeq3}), we get the following
\be
0=(\delta\, \bar\delta-\bar\delta \,\delta)B= \delta (d\bar C)- \bar\delta (d C)=
d (\delta \bar C- \bar \delta C),
\ee
which is nothing but the geometric origin of the CF constraint.

\section{Abelian 3--form gauge theories and 2--gerbes}

In this section, we would like to give a more complex example, 
that of the Abelian 3--form gauge theory, which has not been explicitly worked 
out as yet.
In fact, we derive the proper BRST and anti--BRST symmetry transformations
within a field theoretical approach. In the next section, we will
compare it with the geometric setting of 2--gerbes.

The nilpotent and absolutely anticommuting (anti-)BRST
symmetry transformations for the Abelian 3-form theory 
have been derived \cite{malik4} by exploiting the geometrical superfield
formalism \cite{BT}. Section 4.1 is devoted to the discussion of the
(anti-)BRST invariance of the coupled Lagrangian densities of the theory.
Our Sec.4.2 deals with the discrete and ghost scale symmetry
transformations for the ghost part of the Lagrangian densities.
We obtain the algebraic structures, satisfied by the conserved
charges (corresponding to the above continuous symmetries), in Sec. 4.3.
 
\subsection{Preliminary: nilpotent (anti-)BRST symmetries in
 superfield formulation}

By exploiting the superfield approach to BRST formalism \cite{BT},
the nilpotent and absolutely anticommuting
(anti-)BRST symmetry transformations for the free Abelian
3-form gauge theory have been derived in \cite{malik4}. To be more precise, 
such a theory\footnote{In \cite{malik4}, the theory was considered in 4D, 
but the results hold in
a $D$ dimensional Minkowski space-time. We take here the Greek indices
$\mu, \nu, \eta.......= 0, 1, ..., D-1$ to correspond to the spacetime 
directions of the flat Minkowski spacetime manifold endowed with a metric 
with signatures $(+1, -1, \ldots, -1)$.
For algebraic convenience, we have changed $\bar B \to B,
B_{\mu\nu}^{(2)} \to B_{\mu\nu}, \bar B_{\mu\nu}^{(1)} \to
\bar B_{\mu\nu}$ and exchanged $F_\mu \leftrightarrow \bar F_\mu$
in the notations of \cite{malik4}.}
can be embedded in a $(D, 2)$-dimensional supermanifold
and the application of the celebrated horizontality condition
leads to the following (anti-)BRST symmetry
transformations\footnote{In a field theoretic approach, it is more 
convenient to use the component notation rather than the synthetic differential
form language.}  
\begin{eqnarray}
&&s B_{\mu\nu\eta} = \partial_\mu C_{\nu\eta} + \partial_\nu C_{\eta\mu}
+ \partial_\eta C_{\mu\nu}, \qquad s C_{\mu\nu} = \partial_\mu \beta_\nu
- \partial_\nu \beta_\mu,  \nonumber\\
&&s \bar C_{\mu\nu} = B_{\mu\nu}, \quad
s \bar B_{\mu\nu} = \partial_\mu f_\nu - \partial_\nu f_\mu, \quad
s \bar \beta_\mu = \bar F_\mu, \quad
s \beta_\mu = \partial_\mu C_2,
\nonumber\\
&&s F_\mu = - \partial_\mu B, \quad
s \bar f_\mu = \partial_\mu B_1,
\quad
s \bar C_2 = B_2, \quad s C_1 = - B,  \nonumber\\
&&s \phi_\mu = f_\mu, \quad s \bar C_1 = B_1,
\quad s \Bigl [ C_2, f_\mu, \bar F_\mu, B, B_1, B_2, B_{\mu\nu} \Bigl ] = 0,
\end{eqnarray}
\begin{eqnarray}
&&\bar s B_{\mu\nu\eta} = \partial_\mu \bar C_{\nu\eta} + \partial_\nu \bar C_{\eta\mu}
+ \partial_\eta \bar C_{\mu\nu}, \qquad \bar s \bar C_{\mu\nu} = \partial_\mu \bar \beta_\nu
- \partial_\nu \bar \beta_\mu,  \nonumber\\
&&\bar s  C_{\mu\nu} = \bar B_{\mu\nu}, \quad
\bar s B_{\mu\nu} = \partial_\mu \bar f_\nu - \partial_\nu \bar f_\mu, \quad
\bar s  \beta_\mu =  F_\mu, \quad
\bar s \bar \beta_\mu = \partial_\mu \bar C_2,
\nonumber\\
&&\bar s \bar F_\mu = - \partial_\mu B_2, \quad
\bar s f_\mu = - \partial_\mu B_1,
\quad
\bar s C_2 = B, \quad \bar s C_1 = - B_1,  \nonumber\\
&&\bar s \phi_\mu = \bar f_\mu, \quad \bar s \bar C_1 = - B_2,
\quad \bar s \Bigl [ \bar C_2, \bar f_\mu, F_\mu, B, B_1, B_2, \bar B_{\mu\nu} \Bigl ] = 0,
\end{eqnarray}
where $B_{\mu\nu\eta}$ is the totally antisymmetric tensor gauge field, $(\bar C_{\mu\nu}) C_{\mu\nu}$
are the fermionic antisymmetric (anti-)ghost fields with ghost number $(-1)+1$,
$(\bar\beta_\mu) \beta_\mu$ are the Lorentz vector bosonic ghost-for-ghost
(anti-)ghost fields with ghost
number $(-2)+2$ and $(\bar C_2)C_2$ are the fermionic ghost-for-ghost-for-ghost
(anti-)ghost Lorentz scalar fields with ghost number $(-3)+3$. The vector field $\phi_\mu$ and
auxiliary fields $B, B_1, B_2$ are bosonic in nature and fermionic fields
$(\bar f_\mu)f_\mu$ and $ (\bar F_\mu)F_\mu$ are the auxiliary (anti-)ghost fields with ghost numbers
$(-1)+1$, respectively. Furthermore, it will be noted that the bosonic auxiliary fields
$B$ and $B_2$ carry ghost number $+2$ and $-2$, respectively.

The above off-shell nilpotent ($s_{(a)b}^2 = 0$) (anti-)BRST symmetry transformations
are not absolutely anticommuting in nature because
\begin{equation}
\{s, \bar s  \} B_{\mu\nu\eta} \neq 0, \quad
\{s, \bar s  \} C_{\mu\nu} \neq 0, \quad
\{s, \bar s  \} \bar C_{\mu\nu} \neq 0.
\end{equation}
It is interesting, however, to mention that the superfield formalism \cite{malik4}
yields the following CF type restrictions
\begin{eqnarray}
&& f_\mu + F_\mu = \partial_\mu C_1, \qquad \bar f_\mu + \bar F_\mu = \partial_\mu \bar C_1, \nonumber\\
&& B_{\mu\nu} + \bar B_{\mu\nu} = \partial_\mu \phi_\nu - \partial_\nu \phi_\mu,
\end{eqnarray}
which ensure the absolute anticommutativity (i.e. $s \bar s  + \bar s  s \equiv \{ s, \bar s  \} = 0$)
of the above nilpotent (anti-)BRST symmetry transformations.

We wrap up this section with the following comments:

(i) the nilpotency and absolute anticommutatvity of the (anti-)BRST symmetry transformations are
the indispensable consequences of the superfield approach to BRST formalism \cite{BT},

(ii) there are {\it three} CF type restriction (cf. (21)) for the 4D Abelian 3-form gauge theory whereas
there is one each for the 4D Abelian 2-form \cite{BM} and 4D (non-)Abelian 1-form gauge theories \cite{CF},

(iii) two of the CF type restrictions in (21) are fermionic in nature and one of them is bosonic
(whereas for the Abelian 2-form and (non-)Abelian 1-form gauge theories only bosonic type CF restriction
exists). For the Abelian 1-form gauge theory, the CF type restriction is {\it trivial},

(iv) the CF type restriction is one of the key properties of any arbitrary $p$-form gauge theory
described in the framework of BRST formalism, and

(v) the CF type restrictions, that emerge from superfield formalism,  are always (anti-)BRST invariant
relationships (e.g. $(s, \bar s)\; [f_\mu + F_\mu - \partial_\mu C_1] = 0, \;
(s, \bar s)\; [\bar f_\mu + \bar F_\mu - \partial_\mu \bar C_1] = 0, \;(s, \bar s)\;
[B_{\mu\nu} + \bar B_{\mu\nu} - (\partial_\mu \phi_\nu - \partial_\nu \phi_\mu) ] = 0$).

\subsection{Lagrangian densities: nilpotent (anti-)BRST symmetry transformations}

With the help of the (anti-)BRST symmetry transformations (18) and (19), one can derive the (anti-)BRST
invariant Lagrangian density for the Abelian 3-form gauge theory as
\begin{eqnarray}
{\cal L}_B  &=& \frac{1}{24} H^{\mu\nu\eta\xi} H_{\mu\nu\eta\xi} + s \bar s  \Bigl [ \frac{1}{2}
\bar C_2 C_2 - \frac{1}{2} \bar C_1 C_1 - \frac{1}{2} \bar C_{\mu\nu} C^{\mu\nu} - \bar \beta_\mu \beta^\mu
\nonumber\\
&-& \frac{1}{2} \phi_\mu \phi^\mu - \frac{1}{6} B_{\mu\nu\eta} B^{\mu\nu\eta} \Bigr ], \nonumber\\
{\cal L}_{\bar B}  &=& \frac{1}{24} H^{\mu\nu\eta\xi} H_{\mu\nu\eta\xi} - \bar s  s \Bigl [ \frac{1}{2}
\bar C_2 C_2 - \frac{1}{2} \bar C_1 C_1 - \frac{1}{2} \bar C_{\mu\nu} C^{\mu\nu} - \bar \beta_\mu \beta^\mu
\nonumber\\
&-& \frac{1}{2} \phi_\mu \phi^\mu - \frac{1}{6} B_{\mu\nu\eta} B^{\mu\nu\eta} \Bigr ],
\end{eqnarray}
where $H_{\mu\nu\eta\xi} = \partial_\mu B_{\nu\eta\xi} - \partial_\nu B_{\eta\xi\mu} +
\partial_\eta B_{\xi\mu\nu} - \partial_\xi B_{\mu\nu\eta}$ is the totally antisymmetric
curvature tensor derived from the 4-form $H^{(4)} = d B^{(3)}
\equiv [(dx^\mu \wedge dx^\nu \wedge dx^\eta\wedge dx^\xi)]/ [(4)!] H_{\mu\nu\eta\xi}$. Here
$d = dx^\mu \partial_\mu$ (with $d^2 = 0$) is the exterior derivative and the 3-form
$B^{(3)} = [(dx^\mu \wedge dx^\nu \wedge dx^\eta)]/[(3)!]
B_{\mu\nu\eta}$ defines the totally antisymmetric tensor gauge field $B_{\mu\nu\eta}$. It will be noted
that, within the square brackets of (22), we have taken the combination of
terms that are Lorentz scalars, carry the ghost number
equal to zero and, furthermore, they possess the proper mass dimensions.

Modulo some total spacetime derivatives, the explicit computation of the square
bracketed terms yield the following:
\begin{eqnarray}
&&s \bar s  \Bigl [ \frac{1}{2}
\bar C_2 C_2 - \frac{1}{2} \bar C_1 C_1 - \frac{1}{2} \bar C_{\mu\nu} C^{\mu\nu} - \bar \beta_\mu \beta^\mu
- \frac{1}{2} \phi_\mu \phi^\mu - \frac{1}{6} B_{\mu\nu\eta} B^{\mu\nu\eta} \Bigr ]\nonumber\\
&& = (\partial_\mu B^{\mu\nu\eta}) B_{\nu\eta}  + \frac{1}{2} B_{\mu\nu} \bar B^{\mu\nu}
+ (\partial_\mu \bar C_{\nu\eta} + \partial_\nu \bar C_{\eta\mu} + 
\partial_\eta \bar C_{\mu\nu}) (\partial^\mu C^{\nu\eta})
\nonumber\\
&& - (\partial_\mu \bar \beta_\nu - \partial_\nu \bar \beta_\mu) (\partial^\mu \beta^\nu)
 - B B_2 - \frac{1}{2} B_1^2 + (\partial_\mu \bar C^{\mu\nu}) f_\nu
-  (\partial_\mu  C^{\mu\nu}) \bar F_\nu\nonumber\\
&& + \partial_\mu \bar C_2 \partial^\mu C_2 + \bar f^\mu f_\mu - \bar F^\mu F_\mu + 
(\partial \cdot \beta) B_2 + (\partial \cdot \phi) B_1
- (\partial \cdot \bar \beta) B,
\end{eqnarray}
\begin{eqnarray}
&& - \bar s  s \Bigl [ \frac{1}{2}
\bar C_2 C_2 - \frac{1}{2} \bar C_1 C_1 - \frac{1}{2} \bar C_{\mu\nu} C^{\mu\nu} - \bar \beta_\mu \beta^\mu
- \frac{1}{2} \phi_\mu \phi^\mu - \frac{1}{6} B_{\mu\nu\eta} B^{\mu\nu\eta} \Bigr ]\nonumber\\
&& = - (\partial_\mu B^{\mu\nu\eta}) \bar B_{\nu\eta} + \frac{1}{2} B_{\mu\nu} \bar B^{\mu\nu}
+ (\partial_\mu \bar C_{\nu\eta} + \partial_\nu \bar C_{\eta\mu} + 
\partial_\eta \bar C_{\mu\nu}) (\partial^\mu C^{\nu\eta})
\nonumber\\
&& - (\partial_\mu \bar \beta_\nu - \partial_\nu \bar \beta_\mu) (\partial^\mu \beta^\nu)
 - B B_2 - \frac{1}{2} B_1^2 - (\partial_\mu \bar C^{\mu\nu}) F_\nu
+  (\partial_\mu  C^{\mu\nu}) \bar f_\nu\nonumber\\
&& + \partial_\mu \bar C_2 \partial^\mu C_2 + \bar f^\mu f_\mu - \bar F^\mu F_\mu + 
(\partial \cdot \beta) B_2 + (\partial \cdot \phi) B_1
- (\partial \cdot \bar \beta) B.
\end{eqnarray}
The difference in the above explicit computations is due to the fact that
the (anti-)BRST symmetry transformations are anticommuting only on the
constrained surface defined by the CF type conditions (21).

With the help of the CF type restrictions (21), we can re-express the Lagrangian density
${\cal L}_B$ in an appropriate form as\footnote{In fact, there are, in total, six more possibilities to
express the equivalent Lagrangian densities with the help of CF type restrictions (21). It is, however,
only the equivalent forms like (25) and (26) that have perfect transformations like (27) and (28).}
\begin{eqnarray}
&& {\cal L}_B = \frac{1}{24} H^{\mu\nu\eta\xi} H_{\mu\nu\eta\xi}
+  B^{\mu\nu} \Bigl ( \partial^\eta B_{\eta\mu\nu} + \frac{1}{2} [
\partial_\mu \phi_\nu - \partial_\nu \phi_\mu] \Bigr ) \nonumber\\
&& - \frac{1}{2} B_{\mu\nu}  B^{\mu\nu}
+ (\partial_\mu \bar C_{\nu\eta} + \partial_\nu \bar C_{\eta\mu} + \partial_\eta B_{\mu\nu})
(\partial^\mu C^{\nu\eta}) - (\partial \cdot \bar \beta) B
\nonumber\\
&& - (\partial_\mu \bar \beta_\nu - \partial_\nu \bar \beta_\mu) (\partial^\mu \beta^\nu)
 - B B_2 - \frac{1}{2} B_1^2 + (\partial_\mu \bar C^{\mu\nu} + \partial^\nu \bar C_1) f_\nu
\nonumber\\
&&  -  (\partial_\mu  C^{\mu\nu} - \partial^\nu C_1) \bar F_\nu
+ \partial_\mu \bar C_2 \partial^\mu C_2 + (\partial \cdot \beta) B_2 + (\partial \cdot \phi) B_1.
\end{eqnarray}
It should be noted that, in the above, we have
taken the expression (23) but have replaced $\bar B_{\mu\nu}, \bar f_\mu, F_\mu$
by exploiting the CF type conditions in (21). In an exactly similar fashion, the Lagrangian
density ${\cal L}_{\bar B}$ can be re-written as
\begin{eqnarray}
&& {\cal L}_{\bar B} = \frac{1}{24} H^{\mu\nu\eta\xi} H_{\mu\nu\eta\xi}
- \bar B^{\mu\nu} \Bigl ( \partial^\eta B_{\eta\mu\nu} - \frac{1}{2} [
\partial_\mu \phi_\nu - \partial_\nu \phi_\mu] \Bigr ) \nonumber\\
&& - \frac{1}{2} \bar B_{\mu\nu}  \bar B^{\mu\nu}
+ (\partial_\mu \bar C_{\nu\eta} + \partial_\nu \bar C_{\eta\mu} + \partial_\eta B_{\mu\nu})
(\partial^\mu C^{\nu\eta}) - (\partial \cdot \bar \beta) B
\nonumber\\
&& - (\partial_\mu \bar \beta_\nu - \partial_\nu \bar \beta_\mu) (\partial^\mu \beta^\nu)
 - B B_2 - \frac{1}{2} B_1^2 - (\partial_\mu \bar C^{\mu\nu} + \partial^\nu \bar C_1) F_\nu
\nonumber\\
&&  +  (\partial_\mu  C^{\mu\nu} - \partial^\nu C_1) \bar f_\nu
+ \partial_\mu \bar C_2 \partial^\mu C_2 + (\partial \cdot \beta) B_2 + (\partial \cdot \phi) B_1.
\end{eqnarray}
In the above, it should be noted that we have
taken the expression (24) but have substituted for $B_{\mu\nu}, f_\mu, \bar F_\mu$
by exploiting the CF type conditions in (21).

Under the BRST symmetry transformations (18), the above Lagrangian densities transform 
as 
\begin{eqnarray}
s {\cal L}_B &=& \partial_\mu \Bigl [ (\partial^\mu C^{\nu\eta} + \partial^\nu C^{\eta\mu}
+ \partial^\eta C^{\mu\nu}) B_{\nu\eta} + B^{\mu\nu} f_\nu \nonumber\\
&-& (\partial^\mu \beta^\nu -
\partial^\nu \beta^\mu) \bar F_\nu + B_1 f^\mu - B \bar F^\mu + B_2 \partial^\mu C_2 \Bigr ],
\end{eqnarray}
\begin{eqnarray}
s {\cal L}_{\bar B} &=& - \partial_\mu \Bigl [ (\partial^\mu C^{\nu\eta} + \partial^\nu C^{\eta\mu}
+ \partial^\eta C^{\mu\nu}) \bar B_{\nu\eta} + B^{\mu\nu} F_\nu \nonumber\\
&-& (\partial^\mu \beta^\nu -
\partial^\nu \beta^\mu) \bar f_\nu - B_1 f^\mu + B \bar F^\mu - B_2 \partial^\mu C_2 \nonumber\\
&+& B^{\mu\nu\eta} (\partial_\nu f_\eta - \partial_\eta f_\nu) + \bar C^{\mu\nu} \partial_\nu B
+ C^{\mu\nu} \partial_\nu B_1 \Bigr ] + X,
\end{eqnarray}
where the extra piece $X$, in the above, is given below
\begin{eqnarray}
&& X =  (\partial^\mu C^{\nu\eta} + \partial^\nu C^{\eta\mu}
+ \partial^\eta C^{\mu\nu}) \partial_\mu [ \bar B_{\nu\eta} + B_{\nu\eta} - (\partial_\nu \phi_\eta
- \partial_\eta \phi_\nu) ] \nonumber\\
&& + B^{\mu\nu} \partial_\mu [f_\nu + F_\nu - \partial_\nu C_1]
- [ \bar B_{\mu\nu} + B_{\mu\nu} - (\partial_\mu \phi_\nu
- \partial_\nu \phi_\mu) ] (\partial^\mu f^\nu)
\nonumber\\
&& + (\bar f^\mu + \bar F^\mu - \partial^\mu \bar C_1)
(\partial_\mu B)
- (f^\mu +  F^\mu - \partial^\mu  C_1) (\partial_\mu B_1) \nonumber\\
&& - \;(\partial^\mu \beta^\nu - \partial^\nu \beta^\mu)\;
\partial_\mu [ (\bar f_\nu + \bar F_\nu - \partial_\nu \bar C_1) ].
\end{eqnarray}
Thus, we note that both the above coupled Lagrangian densities are equivalent 
with respect to the BRST transformations (18) on the submanifold defined by the 
CF type field equations (21). In other words, if we impose the CF type restrictions, 
both the above Lagrangian densities respect the BRST symmetry transformations because 
they transform to total derivatives.

In exactly the above fashion, we note that, under the anti-BRST symmetry 
transformations (19), the coupled Lagrangian densities transform as
\begin{eqnarray}
\bar s  {\cal L}_{\bar B} &=& \partial_\mu \Bigl [ (\partial^\mu \bar C^{\nu\eta} + \partial^\nu \bar C^{\eta\mu}
+ \partial^\eta \bar C^{\mu\nu}) \bar B_{\nu\eta} + \bar B^{\mu\nu} \bar f_\nu \nonumber\\
&-& (\partial^\mu \bar \beta^\nu -
\partial^\nu \bar \beta^\mu) F_\nu + B_1 \bar f^\mu + B_2  F^\mu - B \partial^\mu \bar C_2 \Bigr ],
\end{eqnarray}
\begin{eqnarray}
\bar s  {\cal L}_{B} &=&  \partial_\mu \Bigl [ (\partial^\mu \bar C^{\nu\eta} + \partial^\nu \bar C^{\eta\mu}
+ \partial^\eta \bar C^{\mu\nu}) B_{\nu\eta} - \bar B^{\mu\nu} \bar F_\nu \nonumber\\
&+& (\partial^\mu \bar \beta^\nu -
\partial^\nu \bar \beta^\mu)  f_\nu + B_1 \bar f^\mu + B_2  F^\mu - B \partial^\mu \bar C_2 \nonumber\\
&+& (B^{\mu\nu\eta}) (\partial_\nu \bar f_\eta - \partial_\eta \bar f_\nu) + \bar C^{\mu\nu} \partial_\nu B_1
- C^{\mu\nu} \partial_\nu B_2 \Bigr ] + Y,
\end{eqnarray}
where the extra piece $Y$, in the above equation, is
\begin{eqnarray}
&& Y = - (\partial^\mu \bar C^{\nu\eta} + \partial^\nu \bar C^{\eta\mu}
+ \partial^\eta \bar C^{\mu\nu}) \partial_\mu [ \bar B_{\nu\eta} + B_{\nu\eta} - (\partial_\nu \phi_\eta
- \partial_\eta \phi_\nu) ] \nonumber\\
&& + \bar B^{\mu\nu} \partial_\mu [\bar f_\nu + \bar F_\nu - \partial_\nu \bar C_1]
- [ \bar B_{\mu\nu} + B_{\mu\nu} - (\partial_\mu \phi_\nu
- \partial_\nu \phi_\mu) ] (\partial^\mu \bar f^\nu)
\nonumber\\
&&  - (\bar f^\mu + \bar F^\mu - \partial^\mu \bar C_1)
(\partial_\mu B_1)
- (f^\mu +  F^\mu - \partial^\mu  C_1) (\partial_\mu B_2) \nonumber\\
&& - \;(\partial^\mu \bar \beta^\nu
- \partial^\nu \bar \beta^\mu)\;
\partial_\mu [ (f_\nu +  F_\nu - \partial_\nu  C_1) ].
\end{eqnarray}
It is gratifying to note that both the above coupled Lagrangian densities respect the
off-shell nilpotent (anti-)BRST symmetry transformations on the submanifold of the spacetime
that is described by the constrained CF type field conditions (21). These features,
under the (anti-)BRST symmetry transformations, are exactly
like the ones we come across in the context of the BRST
approach to 4D non-Abelian 1-form gauge theory (see, e.g. \cite{NO}).

\subsection{Ghost Lagrangian density: global scale and discrete symmetry transformations}

The ghost parts of the Lagrangian densities (25) and (26), even though 
they look quite different in their appearance,  are
actually {\it equal} on the submanifold defined by
the CF type restrictions (21). Thus, let us take one of them, as 
\begin{eqnarray}
{\cal L}^{(B)}_{(g)} &=& (\partial_\mu \bar C_{\nu\eta} + \partial_\nu \bar C_{\eta\mu} + \partial_\eta
\bar C_{\mu\nu}) (\partial^\mu C^{\nu\eta}) - (\partial \cdot \bar \beta) B 
\nonumber\\
&-& (\partial_\mu \bar \beta_\nu - \partial_\nu \bar \beta_\mu) (\partial^\mu \beta^\nu)
 - B B_2 + (\partial_\mu \bar C^{\mu\nu} + \partial^\nu \bar C_1) f_\nu
\nonumber\\
&-&  (\partial_\mu  C^{\mu\nu} - \partial^\nu C_1) \bar F_\nu
+ \partial_\mu \bar C_2 \partial^\mu C_2 + (\partial \cdot \beta) B_2.
\end{eqnarray}
The above Lagrangian density respects the following continuous global scale 
transformations for the ghost fields, namely;
\begin{eqnarray}
&& C_{\mu\nu} \to e^{+\Omega} C_{\mu\nu}, \qquad \bar C_{\mu\nu} \to e^{-\Omega} \bar C_{\mu\nu}, \qquad
C_{1} \to e^{+ \Omega} C_{1}, \nonumber\\
&&  \bar C_{1} \to e^{- \Omega} \bar C_{1}, \qquad
f_{\mu} \to e^{+ \Omega} f_{\mu}, \qquad
F_{\mu} \to e^{+ \Omega} F_{\mu}, \nonumber\\
&&  \bar f_{\mu} \to e^{-\Omega} \bar f_{\mu}, \qquad
\bar F_{\mu} \to e^{- \Omega} \bar F_{\mu}, \qquad \beta_{\mu} \to e^{+ 2 \Omega} \beta_{\mu}, \nonumber\\
&&\bar \beta_{\mu} \to e^{- 2 \Omega} \bar \beta_{\mu}, \qquad
 B \to e^{+ 2 \Omega} B, \qquad B_2  \to e^{- 2 \Omega} B_2, \nonumber\\
&& C_{2} \to e^{+ 3 \Omega} C_{2}, \qquad
\bar C_{2} \to e^{- 3 \Omega} \bar C_{2},
\end{eqnarray}
where $\Omega$ is a global scale infinitesimal parameter and the numbers,
present in the exponentials, denote the ghost numbers of the corresponding 
dynamical and/or auxiliary (anti-)ghost fields.

In addition to the above global continuous symmetry transformations, the
above Lagrangian density ${\cal L}^{(B)}_g$ also respects the following discrete 
symmetry transformations amongst the (anti-)ghost fields, namely;
\begin{eqnarray}
&& C_{\mu\nu} \to \pm i \bar C_{\mu\nu}, \qquad C_1 \to \mp i \bar C_1, \qquad \beta_\mu \to \pm i \bar \beta_\mu,
\nonumber\\
&&  \bar C_{\mu\nu} \to \pm  i C_{\mu\nu},
\qquad \bar C_1 \to \mp i C_1, \qquad \bar \beta_\mu \to \mp i  \beta_\mu,
\nonumber\\
&& f_\mu \to \pm i \bar F_\mu, \qquad B_2 \to \pm i B, \qquad C_2 \to \pm i \bar C_2, \nonumber\\
&& \bar F_\mu \to \pm i  f_\mu, \qquad B \to \mp i B_2, \qquad \bar C_2 \to \pm i  C_2.
\end{eqnarray}
The above symmetry transformations enable us to go from the BRST symmetry transformations to
the anti-BRST symmetry transformations and {\it vice-versa}. The above transformations have been written for
the ghost part of the Lagrangian density of ${\cal L}_B$. However, similar kind of transformations
can be written out for the ghost part of the Lagrangian density of ${\cal L}_{\bar B}$ where, in
addition to the above transformations, we shall require
\begin{equation}
\bar f_\mu \to \pm i F_\mu, \;\;\qquad \;\;\;F_\mu \to \pm i \bar f_\mu.
\end{equation}
We close this section with the remark that the ghost part of the coupled Lagrangian 
densities remain invariant under (34), (35) and (36). Furthermore, it can be checked 
that the CF type restrictions also remain invariant under the transformations (34), 
(35) and (36) because
$ f_\mu + F_\mu = \pm \partial_\mu C_1$ and $ \bar f_\mu + \bar F_\mu 
= \pm \partial_\mu \bar C_1$
are allowed by the anticommutativity property. It is trivial to state that
$B_{\mu\nu\eta} \to B_{\mu\nu\eta}, B_{\mu\nu} \to B_{\mu\nu}, 
\bar B_{\mu\nu} \to \bar B_{\mu\nu}, \phi_\mu \to \phi_\mu$ under
the ghost transformations because these fields carry ghost number equal to zero.

\subsection{Conserved charges: algebraic structures}

It is clear from equations (27) and (28) that the Lagrangian densities ${\cal L}_B$ 
and ${\cal L}_{\bar B}$ transform precisely to the total spacetime derivative under 
the BRST and anti-BRST transformations (18) and (19). According to the Noether's theorem, 
we have the following expressions for the conserved currents
\begin{eqnarray}
&& J^\mu_{(B)} = s \Phi_i \;
{\displaystyle \frac{\partial {\cal L}_B} {\partial (\partial_\mu \Phi_i)}} - Z^\mu, \nonumber\\
&& J^\mu_{(\bar B)} = \bar s  \Phi_i\;
{\displaystyle \frac{\partial {\cal L}_{\bar B}} {\partial (\partial_\mu \Phi_i)}} - S^\mu,
\end{eqnarray}
where the generic dynamical field $\Phi_i = B_{\mu\nu\eta}, C_{\mu\nu}, 
\bar C_{\mu\nu}, \beta_\mu,
\bar \beta_\mu, \phi_\mu, \bar C_1, C_1, \bar C_2, \bar C_2$ and the explicit expression for
$Z^\mu$ and $S^\mu$ are (cf. (27) and (30))
\begin{eqnarray}
Z^\mu &=&  (\partial^\mu C^{\nu\eta} + \partial^\nu C^{\eta\mu}
+ \partial^\eta C^{\mu\nu}) B_{\nu\eta} + B^{\mu\nu} f_\nu \nonumber\\
&-& (\partial^\mu \beta^\nu -
\partial^\nu \beta^\mu) \bar F_\nu + B_1 f^\mu - B \bar F^\mu + B_2 \partial^\mu C_2,
\nonumber\\
S^\mu &=& (\partial^\mu \bar C^{\nu\eta} + \partial^\nu \bar C^{\eta\mu}
+ \partial^\eta \bar C^{\mu\nu}) \bar B_{\nu\eta} + \bar B^{\mu\nu} \bar f_\nu \nonumber\\
&-& (\partial^\mu \bar \beta^\nu -
\partial^\nu \bar \beta^\mu) F_\nu + B_1 \bar f^\mu + B_2  F^\mu - B \partial^\mu \bar C_2.
\end{eqnarray}
The conserved currents that emerge from the above equations are as follows
\begin{eqnarray}
J^\mu_{(B)} &=& H^{\mu\nu\eta\xi} (\partial_\nu C_{\eta\xi}) + (\partial^\mu C^{\nu\eta} 
+ \partial^\nu C^{\eta\mu} + \partial^\eta C^{\mu\nu}) \; B_{\nu\eta} + B^{\mu\nu} f_\nu 
+ B_1 f^\mu \nonumber\\
&+& B_2 \partial^\mu C_2 - B \bar F^\mu - (\partial^\mu \bar \beta^\nu 
- \partial^\nu \bar \beta^\mu) (\partial_\nu C_2) -  (\partial^\mu \beta^\nu 
- \partial^\nu \beta^\mu) \bar F_\nu \nonumber\\
& -& (\partial^\mu \bar C^{\nu\eta} + \partial^\nu \bar C^{\eta\mu}
+ \partial^\eta \bar C^{\mu\nu}) \; (\partial_\nu \beta_\eta - \partial_\eta \beta_\nu),
\end{eqnarray}
\begin{eqnarray}
J^\mu_{(\bar B)} &=& H^{\mu\nu\eta\xi} (\partial_\nu \bar C_{\eta\xi}) 
+ (\partial^\mu \bar C^{\nu\eta} + \partial^\nu \bar C^{\eta\mu}
+ \partial^\eta \bar C^{\mu\nu}) \; \bar B_{\nu\eta} + \bar B^{\mu\nu} \bar f_\nu 
+ B_1 \bar f^\mu \nonumber\\
&-& B \partial^\mu \bar C_2 + B_2 F^\mu - (\partial^\mu  \beta^\nu 
- \partial^\nu  \beta^\mu) (\partial_\nu \bar C_2)
-  (\partial^\mu \bar \beta^\nu - \partial^\nu \bar \beta^\mu) F_\nu \nonumber\\
&+& (\partial^\mu  C^{\nu\eta} + \partial^\nu  C^{\eta\mu}
+ \partial^\eta  C^{\mu\nu}) \; (\partial_\nu \bar \beta_\eta - \partial_\eta \bar \beta_\nu),
\end{eqnarray}
where all the derivatives of (37) have been calculated from ${\cal L}_B$ and 
${\cal L}_{\bar B}$.

The conservation law (i.e. $\partial_\mu J^\mu_{(i)} = 0, i = B, \bar B$) can be proven 
by taking into account the following equations of motion that are derived from ${\cal L}_B$
\begin{eqnarray}
&& \Box B_{\mu\nu\eta} = 0, \qquad B_{\mu\nu} = (\partial^\eta B_{\eta\mu\nu}) + \frac{1}{2} (\partial_\mu \phi_\nu
- \partial_\nu \phi_\mu), \quad \Box \beta_\mu = 0,  \nonumber\\
&& \Box \bar \beta_\mu = 0,\quad \partial \cdot F = 0, \quad \partial \cdot \bar F = 0, \quad \partial \cdot f = 0,
\quad \partial_\mu B^{\mu\nu} + \partial^\nu B_1 = 0,
\nonumber\\
&& \partial \cdot \bar f = 0, \qquad \partial_\mu C^{\mu\nu} = \partial^\nu C_1, \qquad
\partial_\mu \bar C^{\mu\nu} = - \partial^\nu \bar C_1, \nonumber\\
&& \Box C_1 = 0, \quad \Box \bar C_1 = 0, \quad \Box C_2 = 0, \quad \Box \bar C_2 = 0, \quad \Box B_1 = 0, \nonumber\\
&& B = \partial \cdot \beta, \quad B_1 = \partial \cdot \phi, \quad B_2 = - (\partial \cdot \bar \beta), \quad
\Box \phi_\mu + \partial_\mu (\partial \cdot \phi) = 0,
\nonumber\\
&& \Box C_{\mu\nu} + \frac{1}{2} (\partial_\mu f_\nu - \partial_\nu f_\mu) = 0, \quad
\Box \bar C_{\mu\nu} + \frac{1}{2} (\partial_\mu \bar F_\nu - 
\partial_\nu \bar F_\mu) = 0, \nonumber\\
&& \partial_\mu H^{\mu\nu\eta\xi} + (\partial^\nu B^{\eta\xi} + \partial^\eta B^{\xi\nu} + \partial^\xi B^{\nu\eta}) = 0.
\end{eqnarray}
The Euler-Lagrange equations of motion, that emerge from ${\cal L}_{\bar B}$, are the same as the above but for
the following differences, namely;
\begin{eqnarray}
&& \bar B_{\mu\nu} = - (\partial^\eta B_{\eta\mu\nu}) + \frac{1}{2} (\partial_\mu \phi_\nu - \partial_\nu \phi_\mu),
\quad \partial_\mu \bar B^{\mu\nu} + \partial^\nu B_1 = 0, \nonumber\\
&& \Box C_{\mu\nu} - \frac{1}{2} (\partial_\mu F_\nu - \partial_\nu F_\mu) = 0, \quad
\Box \bar C_{\mu\nu} - \frac{1}{2} (\partial_\mu \bar f_\nu - \partial_\nu \bar f_\mu) = 0, \nonumber\\
&& \partial_\mu H^{\mu\nu\eta\xi} - (\partial^\nu \bar B^{\eta\xi} + \partial^\eta \bar B^{\xi\nu}
+ \partial^\xi \bar B^{\nu\eta}) = 0.  
\end{eqnarray}
It is elementary to state that (41) and (42) are exploited, in a judicious manner,
for the proof of the conservation laws.

The continuous ghost symmetry transformations of (34) lead to the derivation of the ghost conserved
current as given below
\begin{eqnarray}
J^\mu_{(g)} &=& (\partial^\mu \bar C^{\nu\eta} + \partial^\nu \bar C^{\eta\mu}
+ \partial^\eta \bar C^{\mu\nu}) C_{\nu\eta} + (\partial^\mu C^{\nu\eta} + \partial^\nu C^{\eta\mu}
+ \partial^\eta C^{\mu\nu}) \bar C_{\nu\eta}  \nonumber\\
&-& 2 (\partial^\mu \bar \beta^\nu - \partial^\nu \bar \beta^\mu) \beta_\nu  +
2 (\partial^\mu  \beta^\nu - \partial^\nu  \beta^\mu) \bar \beta_\nu + 2 B \bar \beta^\mu + 2 B_2 \beta^\mu
\nonumber\\
&+& 3 (\partial^\mu \bar C_2) C_2 - 3 \bar C_2 \partial^\mu C_2 - \bar C^{\mu\nu} f_\nu - C^{\mu\nu} \bar F_\nu
+ C_1 \bar F^\mu - \bar C_1 f^\mu.
\end{eqnarray}
The conservation law (i.e. $\partial_\mu J^\mu_{(g)} = 0$) can be proven by taking the help of equations of motion
(41) and (42). The conserved (i.e. $\dot Q_{(r)} = 0, r = B, \bar B, g$)
charges (i.e. $Q_{(r)} = \int d^3 x \;J^0_{(r)}, r = B, \bar B, g$), that emerge from
the conserved currents (39), (40) and (43), are as follows
\begin{eqnarray}
Q_{(B)} &=& {\displaystyle \int} d^3 x \Bigl [ H^{0ijk} (\partial_i C_{jk}) +  
(\partial^0 C^{\nu\eta} + \partial^\nu C^{\eta 0}
+ \partial^\eta C^{0 \nu}) B_{\nu\eta} + B_1 f^0 \nonumber\\
&-& (\partial^0 \bar \beta^i - \partial^i \bar \beta^0) \partial_i C_2 - 
(\partial^0  \beta^i - \partial^i  \beta^0) \bar F_i
+ B^{0i} f_i + B_2 \dot C_2 \nonumber\\
&-&  (\partial^0 \bar C^{\nu\eta} + \partial^\nu \bar C^{\eta 0}
+ \partial^\eta \bar C^{0 \nu}) (\partial_\nu \beta_\eta - \partial_\eta \beta_\nu) - 
B \bar F^0
\Bigr ],
\end{eqnarray}
\begin{eqnarray}
Q_{(\bar B)} &=& {\displaystyle \int} d^3 x \Bigl [ H^{0ijk} (\partial_i \bar C_{jk}) - 
 (\partial^0 \bar C^{\nu\eta}
+ \partial^\nu \bar C^{\eta 0} + \partial^\eta \bar C^{0 \nu}) \bar B_{\nu\eta} + 
B_1 \bar f^0 \nonumber\\
&-& (\partial^0  \beta^i - \partial^i \beta^0) \partial_i \bar C_2 - 
(\partial^0  \bar\beta^i - \partial^i  \bar \beta^0) F_i
+ \bar B^{0i} \bar f_i - B \dot {\bar C_2} \nonumber\\
&+&  (\partial^0  C^{\nu\eta} + \partial^\nu  C^{\eta 0}
+ \partial^\eta  C^{0 \nu}) (\partial_\nu \bar \beta_\eta - 
\partial_\eta \bar \beta_\nu) + B_2  F^0
\Bigr ],
\end{eqnarray}
\begin{eqnarray}
Q_{(g)} &=& {\displaystyle \int} d^3 x \Bigl [ 3 \dot {\bar C_2} C_2 - 3 \bar C_2 \dot C_2 
+ (\partial^0 \bar C^{\nu\eta}
+ \partial^\nu \bar C^{\eta 0} + \partial^\eta \bar C^{0 \nu}) C_{\nu\eta} - \bar C_1 f^0
\nonumber\\
&+& 2 (\partial^0 \beta^i - \partial^i \beta^0) \bar \beta_i - 2 (\partial^0 \bar \beta^i - 
\partial^i \bar \beta^0)  \beta_i
- \bar C^{0i} f_i - C^{0i} \bar F_i \nonumber\\
&+& 2 B \bar \beta^0 + 2 B_2 \beta^0 + C_1 \bar F^0  + (\partial^0  C^{\nu\eta}
+ \partial^\nu  C^{\eta 0} + \partial^\eta  C^{0 \nu}) \bar C_{\nu\eta} \Bigr ].
\end{eqnarray}
The above conserved charges are the generators of the nilpotent and
continuous (anti-)BRST as well as continuous ghost scale transformations.

The application of the continuous symmetry transformations on the above charges produces the following algebra
\begin{eqnarray}
&& s Q_{(B)} = - i \{Q_{(B)}, Q_{(B)} \} = 0 \;\;\;\;\;\Rightarrow \;\;Q_{(B)}^2 = 0, \nonumber\\
&& \bar s  Q_{(\bar B)} = - i \{Q_{(\bar B)}, Q_{(\bar B)} \} = 0 \;\;\;\;\Rightarrow
\;\;\; \;Q_{(\bar B)}^2 = 0, \nonumber\\
&& s Q_{(\bar B)} = - i \{Q_{(\bar B)}, Q_{(B)} \} = 0 \;\;\;\;\;\;
\Rightarrow\; \{Q_{(\bar B)}, Q_{(B)} \} = 0, \nonumber\\
&& s Q_{(g)} = - i [Q_{(g)}, Q_{(B)} ] = - Q_{(B)} \;\Rightarrow \;i [Q_{(g)} Q_{(B)} ] = + Q_{(B)}, \nonumber\\
&& \bar s  Q_{(g)} = - i [Q_{(g)}, Q_{(\bar B)} ] = + Q_{(\bar B)} 
\;\;\Rightarrow \;i [Q_{(g)} Q_{(\bar B)} ] = - Q_{(\bar B)}.
\end{eqnarray}
The above algebra is the standard algebra in the BRST formalism. In the above, we have not
written some more transformations on charges. However, those are implied from the above
(e.g. $\bar s  Q_{(B)} = \{ Q_{(B)}, Q_{(\bar B)} \} = 0$, etc.). The above algebra ensures that the ghost
number of the BRST charge is (+1) and that of the anti-BRST charge is (-1).

\section{BRST and anti-BRST for a 2-gerbe field theory}

A 2-gerbe can be described by a quadruple $(C,B,A,f)$, defined by a 3--form $C$, a 2--form B, a 1--form $A$ and a scalar 
$f$ with the following relations. Given a covering $\{U_i\}$ of $M$, we associate
to each $U_i$ a three--form $C_i$. On a double intersection $U_i
\cap U_j$, we have $C_i-C_j = dB_{ij}$. On the triple
intersections $U_i \cap U_j\cap U_k$, we must have $B_{ij}+
B_{jk}+B_{ki}= d A_{ijk}$. On quadruple intersections 
$U_i \cap U_j\cap U_k\cap U_l$, we must have $A_{ijl}-A_{ijk}+A_{jkl}-A_{ikl}=df_{ijkl}$
and on quintuple intersection $U_i \cap U_j\cap U_k\cap U_l\cap U_m$ the following integral cocycle
condition must be satisfied:
\be
f_{ijlm}-f_{ijkm}+f_{ijkl}-f_{iklm}+f_{jklm}= 2\;\pi \;n, 
\qquad n = 0, 1, 2, 3.....\label{3cocycle} 
\ee

Two quadruples $(C,B,A,f)$ and $(C',B',A',f')$ are gauge equivalent if they satisfy
\be
&&B_{ij}'=B_{ij}+\Gamma_i-\Gamma_j+db_{ij},\label{B'B}\\
&&A'_{ijk}= A_{ijk} +b_{ij}+b_{jk}+b_{ki}+d\gamma_{ijk},\label{A'A}\\
&&f'_{ijkl}=  f_{ijkl}+ \gamma_{ijl}-\gamma_{ijk}+\gamma_{jkl}-\gamma_{ikl}.\label{f'f}
\ee

\begin{figure}[htbp]
    \hspace{-0.5cm}
\begin{center}
    \includegraphics[scale=0.5]{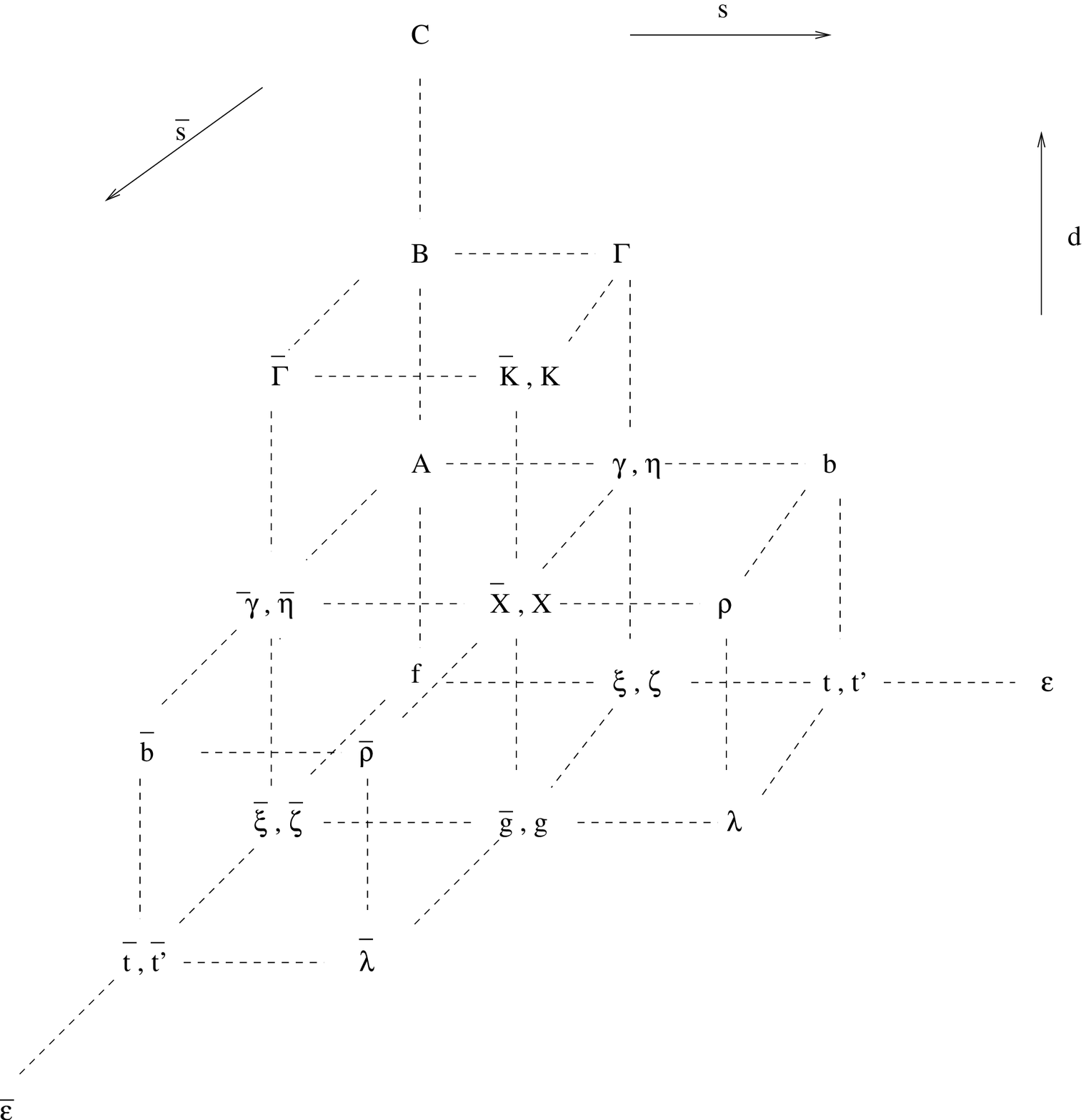}
    \end{center}
\caption{\emph{\small A skematic view of BRST and anti-BRST transformations
for 2--gerbes}}
    \label{fig:B}
\end{figure}

Starting from these transformation properties, we can define the BRST and anti--BRST 
transformations appropriate for the field contents of a 2--gerbe field theory. We have to introduce a 
whole lot of auxiliary fields, even more than in the previous cases, in order
to capture its ample gauge freedom in the language of the absolute anticommutativity
property. The overall field content is summarized 
in Figure 2.

The BRST and anti-BRST transformations, for the boundary faces of the figure, are as follows
\be
&&s\,C=d\Gamma, \qquad s\,\Gamma = db,  \qquad s\, b=d\,\epsilon,\0\\
&& s\, B=\Gamma+d\,\gamma,\qquad s\, \gamma= -b+ d\,t,\qquad s\, t=\epsilon= s\,t',\0\\
&& s\, A = \gamma+\eta+d\,\xi, \qquad s\,\eta= b-d\,t', \qquad s\,\xi=t-t'=-s\, \zeta,
\0\\ && s\, f=\xi+\zeta,\label{2brst1}
\ee
and
\be
&&\bar s\, C=d\bar \Gamma, \qquad\bar  s\,\bar \Gamma = d\bar b,  \qquad \bar s\, 
\bar b=d\,\bar \epsilon,\0\\
&& \bar s\, B=\bar \Gamma+d\,\bar \gamma,\qquad \bar s\, \bar \gamma= -\bar b+ 
d\,\bar t,\qquad \bar s\, \bar t=\bar \epsilon= \bar s\,\bar t',\0\\
&& \bar s\, A = \bar \gamma+\bar \eta+d\,\bar \xi, \qquad \bar s\,\eta= 
\bar b-d\,\bar t', \qquad \bar s\,\bar \xi=\bar t-\bar t '=-\bar s\, 
\bar \zeta,
\0\\ && \bar s\, f=\bar \xi+\bar \zeta,\label{2brst2}
\ee
The remaining transformations are
\be
&&s\, \bar \Gamma = K, \qquad s\,\bar b =-\bar \rho, \qquad \qquad s\,\bar K = d\,\rho,\0\\
&& s\,\bar \gamma= \bar X +d\,\bar g,  \qquad s\,\bar \eta= - \bar X -d\,\bar g,
\qquad s\,g=-\lambda,\0\\
&& s\, X= - \rho,\qquad s\,\bar X= -\,d\lambda, \qquad s\, \bar g=\lambda,\0\\
&& s\,\bar\xi =s\,\bar  \zeta= \frac 12 (g+\bar g),\qquad s\, \bar t= s\, \bar t'=-\bar\lambda,
\label{2brst3}
\ee
and
\be
&&\bar s\, \Gamma =\bar  K, \qquad \bar s\,b =- \rho, \qquad \qquad 
\bar s\, K = d\,\bar\rho,\0\\
&&\bar s  \gamma=  X +d\, g,  \qquad\bar  s\,\eta= - X -d\, g,
\qquad \bar s\,g=-\bar \lambda,\0\\
&&\bar  s\, \bar X= - \bar \rho,\qquad\bar  s\, X= d\bar \lambda, 
\qquad\bar  s\, \bar g=\bar\lambda,\0\\
&& \bar s\,\xi =\bar s\, \zeta=- \frac 12 (g+\bar g),\qquad \bar s\, t=\bar  s\, t'=\lambda.
\label{2brst4}
\ee
Moreover, we also have  $s\,\Big{[}\epsilon,\bar \epsilon, K,\rho, \bar \rho,\lambda,\bar \lambda\Big{]}=0$,
and
$\bar s\,\Big{[}\epsilon,\bar \epsilon,\bar K,\rho, \bar \rho,\lambda,\bar \lambda
\Big{]}=0$,

With these transformation properties, the anticommutator of the BRST and anti-BRST transformations
 $(s\,\bar s+\bar s\,s )$ annihilates all the 
above fields, except for $C$ and $A$, which require the CF condition
\be
K+\bar K+ d(X+\bar X)=0. \label{CF3}
\ee

Once again, we believe, observing the above BRST and anti--BRST algebra, that a geometric 
set-up, similar to the one for 1--gerbes, could be constructed. However, since a 
mathematical formulation
(analogous to Hitchin's  for 1--gerbes) is still missing for 2--gerbes, we leave this 
problem open.

\subsection{Correspondence with the field theory model}

Let us now compare the formulas of this section with the field--theoretical 
derivation of sec. 4. It is clear that in sec. 4 we considered a reduced 
model, where the only non--ghost field is the three--form field.
The correspondence with that Abelian 3--form gauge theory  
can be seen as follows. Suppress the two-form, one--form and 0--form (non--ghost) 
fields of the previous subsection. For the remaining fields the correspondence
is as follows:
\be\label{corresp}
\matrix{ B_{\mu\nu\lambda} \leftrightarrow C, &&
C_{\mu\nu} \leftrightarrow \Gamma&&\bar C_{\mu\nu}, 
\leftrightarrow \bar \Gamma,\cr
\beta_\mu \leftrightarrow b, &&
\bar \beta_\mu \leftrightarrow \bar b, &&
B_{\mu\nu}  \leftrightarrow K,\cr
\bar B_{\mu\nu}  \leftrightarrow \bar K,&&
f_\mu \leftrightarrow \rho,&&
F_\mu \leftrightarrow -\rho,\cr
\bar f_\mu \leftrightarrow \bar \rho,&&
\bar F_\mu \leftrightarrow -\bar\rho,&&
C_2 \leftrightarrow \epsilon,\cr
\bar C_2 \leftrightarrow \bar \epsilon,&&
C_1 \leftrightarrow \lambda,&&
\bar C_1 \leftrightarrow \bar \lambda,\cr
B \leftrightarrow 0,&&
B_1 \leftrightarrow 0,&&
B_2 \leftrightarrow 0}
\ee
and
\be
\phi_{\mu} \leftrightarrow -X-\bar X\label{corresp2}
\ee
The ghost fields $B,B_1,B_2$ are an example of subalgebra that can be moded out.
Upon moding it out, the first two CF constraints in equation (21) 
turn out to be trivial.

\section{Conclusion}

In this paper, we have proposed a geometrical interpretation of the BRST and 
anti-BRST algebra in an ordinary non-Abelian 1-form gauge field theory. To be more 
precise, we
have provided a geometrical interpretation of the CF constraints, which are
needed in order to close the algebra. For this purpose, we have slightly modified
the well-known geometric setting of the gauge theories based on the principal 
fiber bundles to allow for the non-trivial anti-commutator between ghosts and
anti-ghosts fields. 

We have remarked that if the gauge group is Abelian, the CF
conditions are pointless. If an Abelian structure, however, involves a 2-form
gauge field, the non-trivial CF constraints turn up, once again. 
The geometry of such theories is that of 1-gerbes. This geometry dictates
the BRST and anti-BRST structure separately, but a modification is necessary 
in order to interpret the absolute anticommutativity of the joint BRST and anti-BRST
transformations. Based on the previous example of 1-form gauge field theories,
we have suggested how this modification can be implemented.

Finally, we have studied the case of Abelian gauge field theories involving 
3-form fields. We have derived the relevant BRST and anti-BRST 
transformations in the framework of quantum field theory (essentially by using
the superfield method) and, in particular, we have worked out the specific 
CF conditions. The geometry relevant to this kind of theories is based
on the 2-gerbe structure. We have derived the BRST and anti-BRST transformations
from the 2-gerbe geometry and shown that they are compatible with those
obtained via field theory methods. The geometrical interpretation of the
CF conditions, in this context, has still to be spelled out in detail.

\vskip 1cm
\noindent
{\bf Acknowledgements}\\

\noindent
One of us (RPM) would like to acknowledge the financial support
from DST, Govt. of India, under the SERC project grant No: SR/S2/HEP-23/2006.
He would also like to express his deep sense of gratitude to the Director,
AS-ICTP, for the warm hospitality where a part of this work was done.

\end{document}